\documentclass[jkps,fleqn,showpacs,twocolumn,showkeys]{revtex4}
\usepackage{graphicx}
\usepackage{amssymb}
\usepackage{amsmath}
\usepackage{bm}
\usepackage{color}
\usepackage{ulem}

\begin{document}
\setcounter{page}{0}
\title[]{
Thermally-induced magnetic phases in an Ising spin Kondo lattice model \\
on a kagome lattice at 1/3-filling
}
\author{Hiroaki \surname{Ishizuka}}
\email{ishizuka@aion.t.u-tokyo.ac.jp}
\author{Yukitoshi \surname{Motome}}
\affiliation{
Department of Applied Physics, University of Tokyo, 7-3-1 Hongo, Bunkyo, Tokyo
}

\date[]{}

\begin{abstract}
Numerical investigation on the thermodynamic properties of an Ising spin Kondo lattice model on a kagome lattice is reported.
By using Monte Carlo simulation, we investigated the magnetic phases at 1/3-filling.
We identified two successive transitions from high-temperature paramagnetic state to a Kosterlitz-Thouless-like phase in an intermediate temperature range and to a partially disordered phase at a lower temperature.
The partially disordered state is characterized by coexistence of antiferromagnetic hexagons and paramagnetic sites with period $\sqrt3 \times \sqrt3$. 
We compare the results with those for the triangular lattice case. 
\end{abstract}



\maketitle

\section{Introduction}

Geometrical frustration has been one of the major topics in the condensed matter physics for a long time. 
A variaty of peculiar phenomena are emergent from the macroscopic degeneracy of the ground state and its lifting by perturbative interactions~\cite{Diep2005}. 
Thermal or quantum fluctuation also plays a significant role on the system, and in some cases, it gives rise to an exotic phase or unusual order.
A typical example is found in the Ising antiferromagnets on a triangular lattice.
When the interaction is restricted to the nearest neighbor spins, the geometrical frustration prevents the system from long-range ordering, and the system remains disordered down to zero temperature with extensive degeneracy and associated residual entropy~\cite{Wannier1950,Houtappel1950,Husimi1950}.
The extensive degeneracy can be lifted by, e.g., further neighbor interactions and external magnetic field, leading to a variety of magnetic phases.

An example of such unusual phases is the partially disordered (PD) state.
The PD state has a peculiar magnetic order in which paramagnetic sites coexist with magnetically ordered sites.
A possibility of such state was first pointed out in the triangular-lattice Ising model in the presence of second-neighbor ferromagnetic interaction in addition to the nearest-neighbor antiferromagnetic interaction.
Within a mean-field calculation, a three-sublattice PD state was obtained between high-temperature paramagnetic state and low-temperature three-sublattice ferrimagnetic state~\cite{Mekata1977}.
In this PD state, the antiferromagnetically ordered moments on the honeycomb subnetwork coexist with the paramagnetic ones on the remaining sites.
However, subsequent Monte Carlo (MC) calculations have indicated that the PD state is unstable in the purely two-dimensional systems, and is taken over by a Kosterlitz-Thouless (KT) phase~\cite{Takayama1983,Fujiki1984,Fujiki1986}. 
The situation was essentially the same with additional further-neighbor interactions~\cite{Takagi1993} and also on a kagome lattice~\cite{Takagi1995}.
These results imply that the PD state is hard to be stabilized in two-dimensional Ising spin models.

\begin{figure}
\begin{center}
\includegraphics[width=7.8cm]{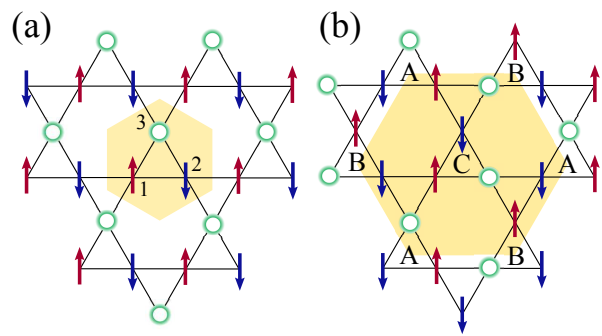}
\end{center}
\caption{
(Color online)
Schematic pictures of the magnetic structure of PD states with (a) ${\bf q}={\bf 0}$ and (b) $\sqrt3\times\sqrt3$ ordering. 
The shaded region in (a) shows the unit cell for the kagome lattice, while the shade in (b) is the magnetic supercell for the $\sqrt3 \times \sqrt3$ order.
The numbers in (a) denote $\nu$ in eq.~(\ref{eq:Sm}), and A,B,C in (b) denote $\alpha$ in eq.~(\ref{eq:Malpha}).
}
\label{fig:pdstate}
\end{figure}

On the other hand, recently, the authors have reported stabilization of a PD state in two dimensions by introducing a coupling to itinerant electrons~\cite{Ishizuka2012}.
By a numerical investigation using MC simulation technique for an Ising spin Kondo lattice model on a triangular lattice, they showed that a three-sublattice PD state appears with opening a (pseudo) charge gap at finite temperatures above a phase-separated region.
The results indicated that the non-perturbative effect of spin-charge coupling plays crucial roles in stabilizing PD.

To address the generality of such PD state stabilized by spin-charge coupling, we here conduct numerical investigation of an Ising spin Kondo lattice model on a kagome lattice.
In the localized spin systems, kagome Ising antiferromagnets have several distinct features compared to the triangular ones. 
When the exchange interaction is restricted to the nearest-neighbor antiferromagnetic one, the kagome model also exhibits macroscopic degeneracy in the ground state.
However, the residual entropy in the kagome case is $S=0.502$~\cite{Kano1953}, which is considerably larger than the value in the triangular case, $S=0.323$~\cite{Wannier1950}.
This is due to the stronger frustration in the kagome lattice consisting of corner-sharing triangles, compared to the edge-sharing triangular lattice; indeed, the spin correlation in the ground state shows an exponential decay in the former, while it is a power-law decay in the latter~\cite{Stephenson1970}.  
For this reason, it is interesting to make a comparative study between the kagome and triangular cases in the presence of the spin-charge coupling.
It will provide a good test for examining the stability of PD state. 

In this contribution, we investigate the thermodynamic behavior of an Ising spin Kondo lattice model on a kagome lattice by using a MC simulation technique. 
As the result, we confirmed two successive transitions: one is from paramagnetic phase to a KT-like magnetic phase, and the other is from the KT-like state to $\sqrt3 \times \sqrt3$ PD state [Fig.~\ref{fig:pdstate}(b)]. 
The results together with the previous ones for the triangular lattice model suggest that the stabilization of PD states by spin-charge coupling is rather general in two-dimensional Ising spin Kondo lattice models.

\section{Model and method}

We consider a single-band Kondo lattice model on a kagome lattice (see Fig.~\ref{fig:pdstate}) with localized Ising spin moments.
The Hamiltonian is given by
\begin{eqnarray}
H = -t \! \sum_{\langle i,j \rangle, \sigma} \! ( c^\dagger_{i\sigma} c_{j\sigma} + \text{H.c.} ) + J \sum_{i}\sigma_i^z S_i.
\label{eq:H}
\end{eqnarray}
The first term represents hopping of itinerant electrons, where $c_{i\sigma}$ ($c^\dagger_{i\sigma}$) is the annihilation (creation) operator of an itinerant electron with spin $\sigma= \uparrow, \downarrow$ at $i$th site, and $t$ is the transfer integral.
The sum $\langle i,j \rangle$ is taken over nearest neighbor sites on the kagome lattice.
The second term is the onsite interaction between localized spins and itinerant electrons, where $\sigma_i^z$ represents the $z$-component of itinerant electron spin and $S_i = \pm 1$ denotes the localized Ising spin at $i$th site; $J$ is the coupling constant (the sign of $J$ does not matter in the present model). 
Hereafter, we take $t=1$ as the unit of energy, the lattice constant $a = 1$, and the Boltzmann constant $k_{\rm B} = 1$.

To investigate thermodynamic properties of this model, we adopted a MC simulation which is widely used for similar models~\cite{Yunoki1998}.
The calculations were conducted up to the system sizes with the number of the unit cell $N=6\times 6$, $6\times 9$, and $9\times 9$, under the periodic boundary conditions 
(the total number of lattice sites is given by $N_{\rm s} = 3N$).
Thermal averages of physical quantities were calculated for typically 4300-9800 MC steps after 1700-5000 MC steps of thermalization. 
In the following, we focus on the electron density at $n=\sum_{i,\sigma} \langle c_{i\sigma}^\dagger c_{i\sigma} \rangle / N_{\rm s} = 1/3$.

To distinguish the PD and other magnetic phases, we use the pseudo-spin~\cite{Fujiki1984} defined for each unit cell 
\begin{eqnarray}
\tilde{\bf S}_{m} = 
\left(
\begin{array}{ccc}
\frac2{\sqrt6} & -\frac1{\sqrt6} & -\frac1{\sqrt6} \\
0              &  \frac1{\sqrt2} & -\frac1{\sqrt2} \\
\frac1{\sqrt3} &  \frac1{\sqrt3} &  \frac1{\sqrt3} \\
\end{array}
\right)
\left(
\begin{array}{c}
S_{m}^{1} \\
S_{m}^{2} \\
S_{m}^{3} \\
\end{array}
\right),
\label{eq:Sm}
\end{eqnarray}
where $S_{m}^{\nu}$ is the Ising spin on $\nu$th site in $m$th unit cell [Fig.~\ref{fig:pdstate}(a)].
Then, we define the sublattice pseudo-spin moment by
\begin{eqnarray}
{\bf M}^{\alpha} = \frac3N \sum_{m \in \alpha} \tilde{\bf S}_{m},
\label{eq:Malpha}
\end{eqnarray}
where $\alpha=\rm A,B,C$ denotes three sublattices for the pseudo-spins [Fig.~\ref{fig:pdstate}(b)]. 
The PD states shown in Fig.~\ref{fig:pdstate} are characterized by a finite ${\bf M}^\alpha = (M_x^\alpha, M_y^\alpha, M_z^\alpha)$ parallel to $(\sqrt{3/2},1/\sqrt2,0)$, $(0,\sqrt2,0)$, or their threefold symmetric directions around the $z$-axis. 
On the other hand, the ferrimagnetic states, in which the paramagnetic spins in the PD states are aligned ferromagnetically, are distinguished by a finite ${\bf M}^\alpha$ along $(\sqrt{2/3},\sqrt2,1/\sqrt3)$, $(2\sqrt{2/3},0,-1/\sqrt3)$, or their threefold symmetric directions around the $z$-axis.
Hence, to parametrize them, we use the azimuth parameter defined by
\begin{eqnarray}
\psi^\alpha = (\tilde{M}_{xy}^\alpha)^3 \cos{6 \phi_{{\bf M}^\alpha}},
\end{eqnarray}
where $\phi_{{\bf M}^\alpha}$ is the azimuth of ${\bf M}^\alpha$ in the $xy$-plane, and $\tilde{M}_{xy}^\alpha = 3 (M_{xy}^\alpha)^2 / 8$ [$M_{xy}^\alpha = \{ (M_x^\alpha)^2 + (M_y^\alpha)^2 \}^{1/2}$]~\cite{Ishizuka2012,Todoroki2004}.
The parameter $\psi^\alpha$ takes a negative value and $\psi^\alpha \to -27/64$ for the ideal PD ordering, while it becomes positive and $\psi^\alpha \to 1$ for the ferrimagnetic ordering; $\psi^\alpha = 0$ for both paramagnetic and KT phases in the thermodynamic limit $N \to \infty$.

For both ${\bf q}={\bf 0}$ and $\sqrt{3} \times \sqrt{3}$ orders shown in Fig.~\ref{fig:pdstate}, $M_{xy}^\alpha$, $|M_z^\alpha|$, and $\psi^\alpha$ are independent of $\alpha$. 
In the following calculations, we found that all the quantities are independent of $\alpha$; hence, we show the averaged values over $\alpha$, $O^\prime = \sum_\alpha O^\alpha /3$ ($O^\alpha = M_{xy}^\alpha, |M_z^\alpha|, \psi^\alpha$). 
To distinguish the ${\bf q}={\bf 0}$ order, we also calculate the net pseudo-spin moment
\begin{eqnarray}
{\bf M} = \frac1N \sum_{m} \tilde{\bf S}_{m} = \frac13 \sum_\alpha {\bf M}^\alpha, 
\end{eqnarray}
and corresponding $M_{xy}$, $|M_z|$, and $\psi$. 
In addition, we compute the susceptibilities for the pseudo-spin moments for each case to detect the phase transitions.

\section{Results and discussion}

\begin{figure}
\begin{center}
\includegraphics[width=8.2cm]{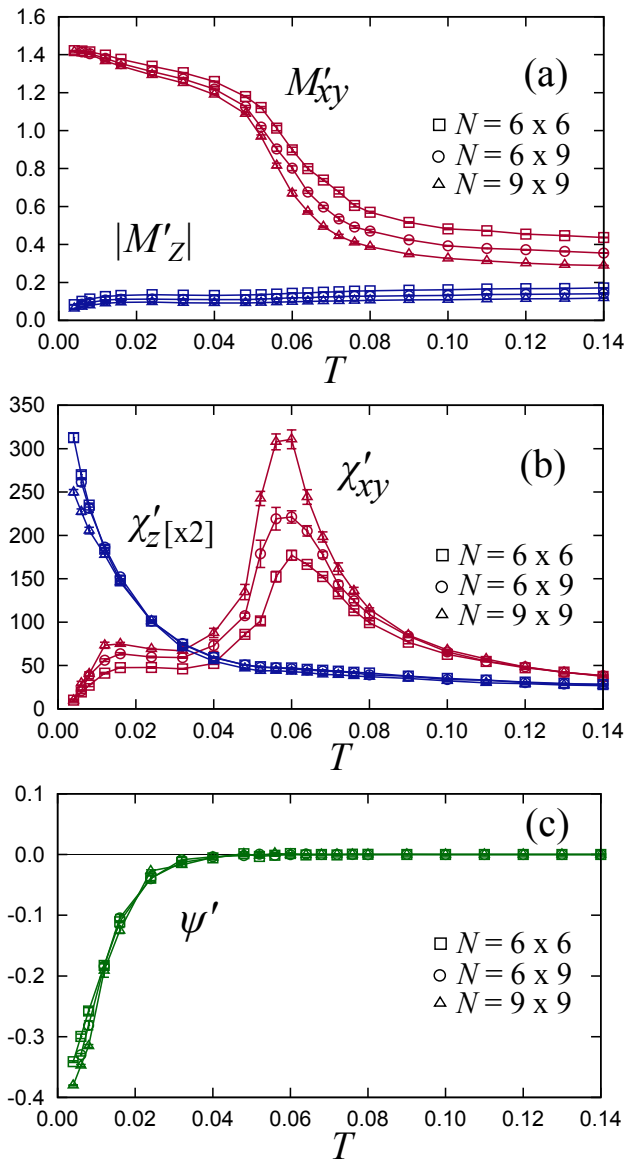}
\end{center}
\caption{
(Color online)
MC results for (a) $M_{xy}^\prime$ and $|M_z^\prime|$, (b) $\chi_{xy}^\prime$ and $\chi_{z}^\prime$, and (c) $\psi^\prime$. 
The data are for $J=2$ and with the system sizes $N= 6 \times 6$, $6\times 9$, and $ 9\times 9$ at $n=1/3$.
}
\label{figure}
\end{figure}

Figure~\ref{figure} shows the results of MC calculation at $n=1/3$ and $J=2$.
As shown in Fig.~\ref{figure}(a), $M_{xy}^\prime$ monotonically increases as decreasing temperature, and shows a rapid increase around $T_{\rm KT}=0.056(3)$.
In addition, it exhibits a small shoulder around $T_c^{\rm PD}= 0.014(2)$ before approaching the value at the lowest temperature.  
The two anomalies are more clearly observed in the corresponding susceptibility $\chi_{xy}^\prime$ plotted in Fig.~\ref{figure}(b); 
$\chi_{xy}^\prime$ shows a divergent peak at $T_{\rm KT}$ and a hump structure at $T_c^{\rm PD}$.
The results imply two successive transitions at $T_{\rm KT}$ and $T_c^{\rm PD}$.

First, we consider the temperature region $T < T_c^{\rm PD}$.
At the lowest temperature of our calculation, $M_{xy}^\prime$ approaches $\sqrt2$, while $|M_{z}^\prime|$ is essentially zero in the thermodynamic limit, as shown in Fig.~\ref{figure}(a).
In addition, $\psi^\prime$ shows a sharp decrease in this region, from $\psi^\prime=0$ to $ \sim - 0.4$, as shown in Fig.~\ref{figure}(c).
The nonzero $\psi^\prime$ indicates a spontaneous breaking of six-fold rotational symmetry of ${\bf M}^\alpha$;
specifically, the negative value approaching $-27/64$ suggests that the system exhibits PD.
Correspondingly, as shown in Fig.~\ref{figure}(b), $\chi_z^\prime$ shows a monotonic increase as decreasing temperature, which is ascribed to the presence of paramagnetic spins in the PD state.
In contrast to $M_{xy}^\prime$, no increase in the net moment $M_{xy}$ nor $M_z$ is seen in the entire range of calculation; $\psi$ is also zero. 
We also found that the spin structure factor exhibits a peak corresponding to the $\sqrt3 \times \sqrt3$ order (not shown).
Hence, we conclude that the system exhibits the PD state with period $\sqrt3 \times \sqrt3$ for $T < T_c^{\rm PD}$.

Next, we examine the intermediate temperature region $T_c^{\rm PD} < T < T_{\rm KT}$.
As shown in Fig.~\ref{figure}(b), $\chi_{xy}^\prime$ shows a divergent peak at $T_{\rm KT}$ corresponding to a rapid rise of $M_{xy}^\prime$ in Fig.~\ref{figure}(a), which is a clear indication of a phase transition.
$M_{xy}^\prime$, however, exhibits
a considerable finite size effect in this temperature region $T_c^{\rm PD} < T < T_{\rm KT}$.
On the other hand, $|M_z^\prime|$ and $\psi^\prime$ shows almost no change. 
In particular, $\psi^\prime$ is extrapolated to zero within statistical errors in the limit of $N \to \infty$, indicating that the intermediate region shows no rotational symmetry breaking in respect to ${\bf M}^\alpha$.  Similar behaviors were observed in the KT phase with quasi long-range order in the Ising antiferromagnets on triangular and kagome lattices~\cite{Takayama1983,Fujiki1984,Fujiki1986,Takagi1993,Takagi1995} and in the Kondo lattice model on a triangular lattice~\cite{Ishizuka2012}. 
Hence, we consider that the intermediate phase for $T_c^{\rm PD} < T < T_{\rm KT}$ is of KT type.

Consequently, our MC data indicate that the system exhibits two successive phase transitions in the calculated temperature range: one is from high-temperature paramagnetic phase to the intermediate KT-like phase, and the other is from the KT-like phase to the low-temperature PD state with period $\sqrt3 \times \sqrt3$.
One point to be addressed here is an anticipated phase transition from the PD state to the true ground state.
The PD state retains residual entropy of $\sim \frac13\log2$ associated with the 1/3 paramagnetic moments. 
Hence, it is unlikely for the PD state to be the ground state of the present model because the degeneracy will be lifted by long-range RKKY interactions induced by the spin-charge coupling~\cite{Ruderman1954,Kasuya1956,Yosida1957}.
However, in our MC simulation at $n=1/3$, there is no indication of further phase transition from PD down to $T \simeq 10^{-3}t$.
This implies that the energy scale of the relevant RKKY interaction is extremely small and that the true ground state is nearly degenerate with the PD state. 
Similar feature was also reported recently in an Ising spin Kondo lattice model on a triangular lattice,
where the PD state survives down to an extremely low temperature at $n=1/3$~\cite{Ishizuka2012}.

Finally, let us compare the present results with those for the model on a triangular lattice~\cite{Ishizuka2012}.
In the kagome lattice case, the transition temperature to the PD state, $T_c^{\rm PD} = 0.014(2)$, is roughly ten times smaller than that in the triangular lattice case, $T_c^{\rm PD} = 0.130(4)$. 
This is presumably because of the stronger frustration in the kagome lattice case. 
Namely, the antiferromagnetic order develops on the honeycomb network in the case of PD on the triangular lattice case, whereas it appears on the disconnected hexagons in the kagome lattice case, as shown in Fig.~\ref{figure}(b); the former can be accommodated by nearest neighbor interactions, but the latter needs further neighbor interactions. 
Hence, the difference of $T_c^{\rm PD}$ can be understood since the RKKY interaction becomes weaker for further neighbors in general. 
In the kagome lattice case, instead, the KT-like phase appears in a wide temperature range above the PD phase, whereas a direct transition from paramagnetic to PD phase was observed for the triangular lattice model at 1/3-filling. 
This might also be ascribed to the difference of frustration; 
higher entropy due to the stronger frustration may contribute to the prevalence of the KT-like phase.

\section{Conclusions}

To summarize, we have investigated the magnetic properties of an Ising spin Kondo lattice model on a kagome lattice at 1/3-filling.
To elucidate the thermodynamic properties of this model, we employed an unbiased Monte Carlo simulation and calculated the pseudo-spin moments, corresponding susceptibilities, and azimuth parameter. 
All the parameters calculated consistently suggested that there are two successive phase transitions, from the high-temperature paramagnetic phase to an intermediate-temperature Kosterlitz-Thouless-like phase, and to the low-temperature $\sqrt3 \times \sqrt3$ partially-disordered state.
Comparison between the present results and those for the triangular lattice case implies that the partial disorder is stabilized in a wide class of two-dimensional Kondo lattice models under frustration.

\begin{acknowledgments}
The authors are grateful to T. Misawa for helpful comments on the calculation.
H.I. is supported by Grant-in-Aid for JSPS Fellows.
This research was supported by KAKENHI (No. 19052008, 21340090, 22540372, and 24340076), Global COE Program
``the Physical Sciences Frontier", the Strategic Programs for Innovative Research (SPIRE), MEXT, and the
Computational Materials Science Initiative (CMSI), Japan.
\end{acknowledgments}

\end{document}